\documentstyle[preprint,aps]{revtex}

\begin{document}

\draft         % PRINTS PACS NUMBER

\preprint{\font\fortssbx=cmssbx10 scaled \magstep2
\hbox to \hsize{
%\special{psfile=uwlogo.ps
%hscale=8000 vscale=8000
%hoffset=-12 voffset=-2}
\hskip.5in \raise.1in\hbox{\fortssbx University of Wisconsin - Madison}
\hfill$\vcenter{\hbox{\bf MADPH-95-912}
                \hbox{\bf IFUSP-P 1135}
                \hbox{\bf hep-th/9510063}
                \hbox{October 1995}
}$}}

\title{ Quantum evolution of scalar fields \\
in Robertson-Walker space-time}

\author{H.\ C.\ Reis \cite{hcr} }

\address{Instituto de F\'{\i}sica,
Universidade de S\~ao Paulo, \\
C.P.\ 66318, 05389-970 S\~ao Paulo, Brazil}

\author{O.\ J.\ P.\ \'Eboli \cite{ojpe} }

\address{Physics Department, University of Wisconsin, \\
Madison, WI 53706, USA}

\date{October 3, 1995}
\maketitle

\begin{abstract}

  We study the $\lambda \phi^4$ field theory in a flat
  Robertson-Walker space-time using the functional Sch\"odinger
  picture.  We introduce a simple Gaussian approximation to analyze
  the time evolution of pure states and we establish the
  renormalizability of the approximation. We also show that the
  energy-momentum tensor in this approximation is finite once we
  consider the usual mass and coupling constant renormalizations.

\end{abstract}

\vskip 1.in

\begin{center}
{\em Submitted to Physical Review D15}
\end{center}

\newpage

%**********
% PAPER
%**********
\section{Introduction}
\label{sec:int}

Establishing the validity of various cosmological scenarios requires
understanding the dynamical evolution of the cosmos in time, {\em e.\
g.\/}, the time evolution of the inflation-driving field \cite{inf:1}
before, during and after inflation, and the details of symmetry changing
phase transitions that may have given rise to cosmic strings \cite{str}.
The conventional formulation of quantum field theory in terms of causal
Green's functions in the Heisenberg picture is not especially suited to
time-dependent problems that make use of an initial condition for
specific solution. Green's functions contain all information needed for
determining transition rates, $S$-matrix elements, etc., of systems in
equilibrium where initial data are superfluous. However, following a
system's time evolution from a definite initial configuration is more
efficiently accomplished in a Schr\"odinger picture description, where
the initial data consist of specifying a pure or mixed state.

For bosonic fields, the field-theoretic functional Schr\"odinger
picture \cite{sch} is a generalization from ordinary quantum mechanics
to the infinite number of degrees of freedom that constitute a field.
Therefore, it allows the use of the mathematical/physical intuition
acquired in quantum mechanics to analyze field-theoretic problems.
Notwithstanding, the functional Schr\"odinger picture is not as widely
used in actual calculations as the Green's function method since
renormalization is more easily carried out in the latter framework.
However, it has been established renormalizability for of the
Schr\"odinger picture for both static \cite{r:sta} and time-dependent
\cite{r:td} cases.

In the Schr\"odinger picture, a pure state is given by a single wave
functional $\Psi(\phi)$ of a $c$-number field $\phi({\bf x})$ at
fixed time, while a pure state is described by a functional density
matrix $ \rho ( \phi_1, \phi_2 ) = \sum_n p_n \Psi_n(\phi_1)
\Psi_n^*(\phi_2)$, where $\{ \Psi_n\} $ is a complete set of wave
functionals and $p_n$ is the probability that the system is in the
state $n$ \cite{ejp}. The time evolution of pure and mixed is governed
by the time-dependent Schr\"odinger equation and the Liouville-von
Neumann equation, respectively. In general these equations cannot be
solved, except for systems whose Hamiltonian is quadratic, and
consequently approximation methods are needed.  Variational
approximations were developed in Refs.\ \cite{ejp,jk,bv} which lead to
tractable equations in the cases of Gaussian trial states. The
Gaussian approximation leads to self-consistent equations that, unlike
perturbation theory, reflect some of the non-linearities of the full
quantum theory.

In various inflationary models, the dynamics of the early universe is
dictated by the time evolution of an scalar field, usually called
inflaton. The equation of motion used in most of the analyses is of the
form
\begin{equation}
\ddot{\varphi} + 3 \frac{\dot{a}}{a} \dot{\varphi} +
V_{\text{eff}}^\prime = 0 \; ,
\end{equation}
where $V_{\text{eff}}$ is the static, finite-temperature effective
potential. However, the static effective potential does not properly
take into account effects of non-equilibrium dynamics. Moreover,
semiclassical \cite{semi} or linearized approximations \cite{line}
have been frequently made, and the full non-linearity of an
interacting theory is lost. It is conceivable that a more complete
analysis may change the picture of the early universe drawn from these
approximations \cite{sep,boya}.

In this work we study a self-interacting scalar quantum field model in
a flat Robertson-Walker space-time in the functional Schr\"odinger
picture. Our purpose is to obtain a set of quantum dynamical equations
describing the evolution of a scalar field as well as the scale
factor. In order to obtain a small set of tractable equations, which
capture some of the non-linearities of the full problem, we employ a
variational method using a Gaussian trial state whose kernels depend
on just a few parameters. We analyze the renormalization of the
equations of motion for the variational parameters to assure the
consistency of them. Moreover, we also study the renormalization of
the expectation value of the energy-momentum tensor, which is the
source for the semi-classical Einstein equation, showing that it is
finite once we take into account the usual mass and coupling constant
renormalizations.  Therefore, we obtain a set of equations that can be
used to address various cosmological questions like the dynamics of
chaotic inflation.

This article is organized as follows: In Sec.\ II we present the
functional Schr\"odinger picture for a scalar field and obtain the
Gaussian approximation for the ground state of the $\lambda \Phi^4$
model, which will be used to define our simplified Gaussian
approximation. Section III contains the {\em Ansatz} that we used in
our calculations as well as the study of the renormalization of the
equations of motion for the variational parameters. We present the
analyses of the renormalization of the energy-momentum tensor in Sec.\
IV and, in Sec.\ V we draw our conclusions.

%%%%%%%%%%%

\section{Functional Schr\"odinger picture}

\subsection{Generalities}

Here we present the basic facts about the functional Schr\"odinger
picture, and the interested reader can find explicit examples and
learn more about it in Refs.\ \cite{sch,ejp,sep}. In the
field-theoretic Schr\"odinger picture, pure states are described by
wave functionals $\Psi(\phi)$ of a $c$-number $\phi({\bf x})$ at a
fixed time. The inner product is defined by functional integration
\begin{equation}
\langle \Psi_1 | \Psi_2 \rangle \Rightarrow \int D\phi~ \Psi_1^*
(\phi) \Psi_2(\phi) \; ,
\end{equation}
while operators are represented by functional kernels.
\begin{equation}
O | \Psi \rangle \Rightarrow \int D\phi^\prime~ O(\phi,
\phi^\prime) \Psi(\phi^\prime)
\end{equation}

We adopt a diagonal kernel $\Phi({\bf x}) \Rightarrow \phi({\bf x)}
\delta(\phi - \phi^\prime)$ for the canonical field operator
$\Phi({\bf x})$ at fixed time, while the canonical commutation relations
determine the canonical momentum kernel to be $\Pi({\bf x}) \Rightarrow
(1/i) [\delta/\delta\phi({\bf x})] \delta(\phi - \phi^\prime)$.
Hence $\Phi$ acts by multiplication on functionals of $\phi$ and
$\Pi$ acts by functional differentiation. In this way, the action of any
operator constructed from $\Pi$ and $\Phi$ is
\begin{equation}
{\cal O}(\Pi, \Phi) | \Psi \rangle \Rightarrow
{\cal O}\left [ \frac{1}{i} \frac{\delta}{\delta \phi}, \phi
\right ] \Psi(\phi) \; .
\end{equation}

The fundamental equation for initial value problems is the
time-dependent Schr\"odinger equation for the time-dependent
state functional $\Psi( \phi; t)$. This equation takes definite form,
once a Hamiltonian operator $H(\Pi, \Phi)$ is specified:
\begin{equation}
i \frac{\partial}{\partial t} \Psi(\phi; t ) =
H \left [  \frac{1}{i} \frac{\delta}{\delta \phi}, \phi
\right ] \Psi(\phi; t) \; .
\label{eq:schr}
\end{equation}
The initial value problem is completely defined once we also supply
the initial wave functional.

\subsection{Dirac's Variational Principle}

The time-dependent Schr\"odinger equation (\ref{eq:schr}) cannot be
directly integrated, unless the system is described by a quadratic
Hamiltonian. Therefore, we shall employ a variational approximation to
study non-linear (interacting) systems. Applications of variational
principles with restricted variational {\em Ansatz}, in the
Rayleigh-Ritz manner, result in tractable self-consistent dynamical
equations for the parameters used in the {\em Ansatz}, which still
retain some of the non-linearity of the complete problem.

For pure states, the time-dependent Schr\"odinger equation can be obtained
through Dirac's variational principle \cite{dirac}. First we define
the effective action $\Gamma$ as the time integral of the diagonal
matrix element of $i\partial/\partial t -H$:
\begin{equation}
\Gamma = \int dt~ \left \langle \Psi \left | i \frac{\partial}{\partial t}
-H \right | \Psi \right \rangle \; ,
\label{dirac}
\end{equation}
and then we demand that $\Gamma$ be stationary against arbitrary
variations of $|\Psi \rangle$ and $\langle \Psi |$, imposing
appropriate boundary conditions.

It is possible to implement Dirac's principle in two steps in such way
that $\Gamma$ can be associated to the functional generator of the
one-particle irreducible Green's functions with arbitrary energy and
momentum \cite{jk}. In order to do so, we consider the time integral
of an off-diagonal matrix element
\begin{equation}
\Gamma = \int dt~  \left \langle  \Psi_- \left | i \frac{\partial}{\partial
t} - H \right | \Psi_+ \right \rangle \; ,
\end{equation}
subject to the constraint that the matrix element of the field $\Phi
({\bf x})$  is held fixed at a prescribed function $\varphi({\bf x}, t)$.
\begin{eqnarray}
\langle \Psi_- | \Phi({\bf x}) | \Psi_+ \rangle &=& \varphi({\bf x}, t)
\\
\langle \Psi_- | \Psi_+ \rangle &=& 1
\end{eqnarray}
These constraints are supplemented by the boundary condition that these
states tend to the ground state of $H$ for $t \rightarrow \pm \infty$
The physical theory is recovered when we remove the constraints by
solving
\begin{equation}
\frac{\delta \Gamma}{\delta  \varphi({\bf x}, t)} = 0 \; .
\end{equation}

\subsection{Gaussian {\em Ansatz} }

Dirac's variational principle can be used to obtain approximations
provided we restrict the variation of the trial wave functional $\Psi$
to a subspace of the full Hilbert space. In this work, we shall use
Gaussian trial states, whose most general expression is
\begin{eqnarray}
\Psi( \phi, t) = && N(t) ~ \exp \left \{ i \int_{\bf x} \hat{\pi}({\bf x},t)
\left [  \phi ({\bf x})  - \varphi ({\bf x},t)
\right ] \right \}
\nonumber \\
&& \times \exp \left \{ -\int_{{\bf x},{\bf y}}
[\phi({\bf x})-\varphi({\bf x},t)]
\left[ \frac{1}{4}\Omega^{-1}({\bf x},{\bf y},t)-i\Sigma
({\bf x}, {\bf y},t) \right]
\left[ \phi({\bf y})-\varphi({\bf y},t) \right ]
\right \} \; ,
\label{gauss:pure}
\end{eqnarray}
where the variational parameters are $\varphi$, $\hat\pi$, $\Omega$,
and $\Sigma$ and we abbreviated the integral in $d$ spatial dimensions
as $\int_{\bf x} \equiv \int d^d{\bf x}$.  The physical meaning of the
parameters of this wave functional can be inferred from linear and
bilinear averages.  The linear averages are given by
\begin{eqnarray}
\langle \Phi({\bf x}) \rangle &=& \varphi({\bf x}, t)
\; , \\
\langle \Pi({\bf x}) \rangle &=& \hat{\pi}({\bf x}, t)
\; ,
\end{eqnarray}
while bilinear averages are
\begin{eqnarray}
\langle \Phi({\bf x}) \Phi({\bf y}) \rangle &=& \varphi({\bf x})
\varphi({\bf y}) + \Omega ({\bf x}, {\bf y}, t)
\; , \\
\langle \Pi({\bf x}) \Pi({\bf y}) \rangle &=& \hat\pi({\bf x})
\hat\pi({\bf y}) + \frac{1}{4}
\Omega^{-1} ({\bf x}, {\bf y}, t)
+~4 \left(\Sigma \Omega
\Sigma \right)({\bf x}, {\bf y}, t)
\; , \\
\langle \Phi({\bf x}) \Pi({\bf y}) \rangle &=& \frac{i}{2}
\delta({\bf x}-{\bf y}) + 2 \left (
\Omega \Sigma \right )   ({\bf x}, {\bf y}, t)
\; ,
\end{eqnarray}
where we have used the matrix notation $ \left({\cal O} {\cal K}\right)({\bf
x},
{\bf y}) = \int_{{\bf z}} {\cal O}({\bf x}, {\bf z}) {\cal K}({\bf z},
{\bf y})$.  Moreover, from the average value of the operator $i (
\partial/\partial t)$ appearing in the effective action
(\ref{dirac}), we find that the imaginary part of the covariance
function ($\Sigma$) plays the r\^ole of a canonical momentum conjugate
to the real part $\Omega$.
\begin{equation}
\left \langle i \frac{\partial}{\partial t} \right \rangle =
\int_{\bf x} \hat\pi({\bf x}, t) \dot\varphi({\bf x}, t)
+ \int_{\bf x,y} \Sigma({\bf x}, {\bf y}, t) \dot\Omega
({\bf y}, {\bf x}, t)
\end{equation}

\subsection{Gaussian Vacuum in Flat Space-Time}

In order to gain some intuition and also to motivate the
simplification of the Gaussian {\em Ansatz}, that we shall use in this
work, it is interesting to obtain the vacuum state, in the Gaussian
approximation, for the $\lambda \Phi^4$ model in Minkowski space-time.
The dynamics of this system is governed by the Hamiltonian
\begin{equation}
H =\int_{\bf x} \left\{\frac{1}{2}[ \Pi^{2}
+ ({\bf \nabla}\Phi)^{2}]+U(\Phi)\right\} \;,
\end{equation}
where the potential function $U(\Phi)$ is
\begin{equation}
U(\Phi)=\frac{\mu^2}{2} \Phi^{2} +\frac{\lambda}{4!}{\Phi}^{4} \; .
\label{pot:u}
\end{equation}

Substituting the wave functional (\ref{gauss:pure}) into the Dirac's
variational principle leads to
\begin{eqnarray}
\Gamma(\varphi,{\hat{\pi}},\Omega,{\Sigma})&=&\int{dt}\int_{{\bf x}}
    \left\{\left[\hat{\pi}\dot{\varphi}- \left(\frac{1}{2}
    \hat{\pi}^{2}+\frac{1}{2} ({\bf \nabla}\varphi)^{2}
    +U(\varphi)\right)\right]\right.\nonumber\\
&+& \hbar \left[\left(\Sigma{\dot{\Omega}}\right)({\bf x},{\bf x},t)
    -\frac{1}{8}\Omega^{-1}({\bf x},{\bf x},t)-2
    \left(\Sigma{\Omega}\Sigma\right)({\bf x},{\bf x},t) \right. \nonumber\\
&-& \left. \frac{1}{2}(- {\bf \nabla}_{{\bf x}}^{2}\Omega({\bf x},
    {\bf y},t)\mid_{{\bf x}={\bf y}}+U^{(2)}(\varphi)\Omega
({\bf x}, {\bf x}, t )\right]\nonumber\\
&-& \left. \frac{\hbar^{2}}{8}U^{(4)}(\varphi)
    \Omega({\bf x},{\bf x},t)\Omega({\bf x},{\bf x},t)\right\} \;\;,
\label{flat:act}
\end{eqnarray}
where $U^{(n)} \equiv d^nU/d\varphi^n$. The terms in the first square
bracket are the classical action, while terms in the second (last)
square bracket are formally ${\cal O}(\hbar)$ (${\cal O}(\hbar^2)$)
corrections.  In fact, the expression (\ref{flat:act}) contains all
powers in $\hbar$ since the kernel $\Omega$ must satisfy a
self-consistent equation, as shown below.  By varying (\ref{flat:act})
with respect the parameters in our {\em Ansatz} we obtain four
variational equations
\begin{eqnarray}
\frac{\delta{\Gamma}}{\delta{\varphi}({\bf x},t)}=0 &\Longrightarrow&
 \hat{\pi} ({\bf x},t) = \dot{\varphi}({\bf x},t)
\; ,
\label{eq:gen:1}\\
\frac{\delta{\Gamma}}{\delta\hat{\pi}({\bf x},t)}=0 &\Longrightarrow &
\dot{\hat{\pi}}= \left[ {\bf \nabla}_{{\bf x}}^{2}-U^{(1)}(\varphi)
-\frac{1}{2}U^{(3)}(\varphi)\Omega({\bf x}, {\bf x}, t)\right]\varphi
\; ,  \\
\frac{\delta{\Gamma}}{\delta{\Sigma}({\bf x},{\bf y},t)}=0 &\Longrightarrow&
\dot{\Omega}({\bf x},{\bf y},t)= 2 [(\Omega{\Sigma})({\bf x},{\bf y},t)
+(\Sigma{\Omega})({\bf x},{\bf y},t)]
\; , \\
\frac{\delta{\Gamma}}{\delta{\Omega}({\bf x},{\bf y},t)}=0 &\Longrightarrow&
\dot{\Sigma}({\bf x},{\bf y},t)= \frac{1}{8}
\Omega^{-2}({\bf x},{\bf y},t)-2 \Sigma^{2}({\bf x},{\bf y},t)
\nonumber \\
& &- \left. \frac{1}{2}\left[- {\bf \nabla}_{{\bf x}}^{2}+U^{(2)}(\varphi)
+\frac{1}{2}U^{(4)}(\varphi)\Omega({\bf x},{\bf x},t)\right]
\delta ({\bf x}-{\bf y})\right\}
\; .
\label{eq:gen:2}
\end{eqnarray}

Translation invariance implies that $\varphi$ is homogeneous and that
the kernels can be expressed as a Fourier transformation (FT)
\begin{equation}
\Omega ({\bf x}, {\bf y}, t) = \int_{{\bf k}} e^{i{\bf k} \cdot ({\bf x}
- {\bf y})} \Omega({\bf k}, t) \; ,
\end{equation}
where the momentum-space integral $(\int~d^d{\bf k}/(2\pi)^d)$ is
denoted by $\int_{\bf k}$.  Moreover, the kernels for the vacuum state
are time-independent and the above equations of motion reduce to
\begin{eqnarray}
\hat\pi &=& 0
\; ,
\label{pi:0}\\
\varphi \left (\mu^2 + \frac{\lambda}{6} \varphi^2 + \frac{\lambda}{2}
\int_{\bf k} \Omega({\bf k}) \right ) &=& 0
\; , \\
\Omega({\bf k}) \Sigma({\bf k}) &=& 0 \; ,
\\
 \frac{1}{8} \Omega^{-2}({\bf k}) - 2 \Sigma^2({\bf k})
-\frac{1}{2} \left ( {\bf k}^2 + \mu^2 + \frac{\lambda}{2} \varphi^2
+ \frac{\lambda}{2} \int_{\bf k} \Omega({\bf k}) \right )
&=& 0 \; ,
\end{eqnarray}
whose solution is $\hat\pi = \Sigma({\bf k}) = 0$ and
\begin{equation}
\Omega({\bf k}) = \frac{1}{2 \sqrt{ {\bf k}^2 + m^2 } } \; ,
\label{stat}
\end{equation}
with $m^2$ satisfying the gap equation
\begin{equation}
m^2 = \mu^2 + \frac{\lambda}{2} \varphi^2 + \frac{\lambda}{2}
\int_{\bf k} \Omega({\bf k})
\label{stat:m}\; .
\end{equation}
Since this last equation is a self-consistent one for $m^2$, some of the
non-linearities of the complete problem are retained by the Gaussian
approximation.

\subsection{Renormalization of the Effective Potential}

At this point it is instructive to study the renormalization of
the Gaussian effective potential, which can be obtained from the
effective action through
\begin{equation}
\left. \Gamma(\varphi,\hat{\pi},\Sigma,\Omega)\right
|_{\mbox{\scriptsize {static}}}
=-V_{\text{eff}}(\varphi,\hat{\pi},\Sigma,\Omega)\int_{{\bf x}} \; .
\end{equation}
Therefore, from Eq.\ (\ref{flat:act}) we can see that the Gaussian
effective potential is
\begin{eqnarray}
V_{\text{eff}}(\varphi,\hat{\pi},\Sigma,\Omega)
                &=& \frac{1}{2}\hat{\pi}^{2}+\frac{\mu^{2}}{2}\varphi^{2}
                    +\frac{\lambda}{4!} \varphi^{4}
                    \nonumber\\
                &+& \frac{1}{8}\Omega^{-1}({\bf x},{\bf x})+2\left(\Sigma
                    \Omega \Sigma \right)({\bf x},{\bf x})-\frac{1}{2}
                    {\bf \nabla}_{{\bf x}}^{2} \left. \Omega({\bf x},
                    {\bf y})\right |_{{\bf x}={\bf y}} \nonumber\\
                &+& \frac{1}{2}\left(\mu^{2}+\frac{\lambda}{2} \varphi^{2}
                    \right)\Omega({\bf x},{\bf x})+\frac{\lambda}{8}
                    \Omega({\bf x},{\bf x})\Omega({\bf x},{\bf x}) \; .
\end{eqnarray}

The effective potential $V_{\text{eff}}(\varphi)$ is obtained from
$V_{\text{eff}}(\varphi,\hat{\pi},\Sigma,\Omega)$ by minimizing with
respect to the parameters $\hat\pi$, $\Omega$, and $\Sigma$. This
procedure leads to $\hat\pi = \Sigma({\bf k}) = 0$ and Eqs.\
(\ref{stat}) and (\ref{stat:m}), resulting in
\begin{eqnarray}
V_{\text{eff}}(\varphi) &=& \frac{\mu^{2}}{2}\varphi^{2}+\frac{\lambda}{4!}
                            {\varphi}^{4}+\frac{1}{4}\int_{{\bf k}}\sqrt{k^{2}
                            +m^2} \nonumber\\
                        &+& \frac{1}{4}\int_{{\bf k}}\left(k^{2}+\mu^{2}
                            +\frac{\lambda}{2} {\varphi}^{2}\right)
                            \frac{1}{\sqrt{k^{2}+m^2}} \nonumber\\
                        &+& \frac{\lambda}{32} \int_{{\bf k},{\bf k}'}
                            \frac{1}{\sqrt{k^{2}+m^2}}
                            \frac{1}{\sqrt{k^{'2}+m^2}} \;\;.
\end{eqnarray}

In the limit $d=3$, the above integrals are clearly divergent, due the
short distance behavior of $\Omega$, and we must renormalize
$V_{\text{eff}}$.  We regularized these integrals by using dimensional
regularization \cite{jj} on the spatial dimension $d$, obtaining that
the regularized effective potential is
\begin{eqnarray}
V_{\text{eff}}(\varphi)&=&\frac{\mu^2}{2} \varphi^{2}+\frac{\lambda}{4!}
{\varphi}^{4}
+\frac{1}{(4\pi)^{(d+1)/2}(1-d)}\left(\frac{m^2}{\Lambda^{2}}\right)
^{(d-3)/2}\Gamma \left(\frac{3-d}{2}\right)
\nonumber\\
 &\times& \left\{\frac{2m^4}{(1+d)}-m^4+\left(\mu^{2}+
\frac{\lambda}{2} {\varphi^{2}}\right)m^2
\right. \nonumber\\
 &+& \left. \frac{\lambda}{2}{m^4}\frac{1}{(4\pi)^{(d+1)/2}(1-d)}
\left(\frac{m^2}{\Lambda^{2}}\right)^{(d-3)/2}\Gamma \left(\frac{3-d}{2}
\right)\right\} \; ,
\label{veff:0}
\end{eqnarray}
where $\Lambda$ is an arbitrary mass scale. Notice that the term
$\lambda \Omega \Omega$ gives rise to a double pole in $d=3$, while
the other divergences are single poles. Since the Gaussian
approximation is very similar to the large-$N$ approximation, the
effective potential becomes finite by the renormalization prescription
\cite{presc}
\begin{eqnarray}
\frac{\mu^{2}}{\lambda}&=& \frac{\mu^{2}_{R}}{\lambda_{R}} \; ,
\label{mu:r} \\
\frac{1}{\lambda} &=& \frac{1}{\lambda_R} -
\frac{2}{(4\pi)^{(d+1)/2} }~ \frac{1}{(d-1)(3-d)} \; .
\label{lambda:r}
\end{eqnarray}
This last relation implies that
\begin{equation}
\lim_{d \to 3}\lambda = -16\pi^{2}(3-d)\left\{1+\frac{16\pi^{2}(3-d)}
{\lambda_{R}}+O[(3-d)^{2}]\right\} \text{  for  } \lambda_{R}\neq 0 \; .
\end{equation}

When the above renormalization prescription is substituted into Eq.\
(\ref{veff:0}), it leads to a finite expression for the effective
potential
\begin{eqnarray}
V_{\text{eff}}(\varphi)=\frac{m^2}{2}\varphi^{2} - \frac{m^4}{64\pi^{2}}
{}~\gamma+\frac{m^4}{64\pi^{2}}{\ln}\left(\frac{m^2}
{4\pi\Lambda^{2}}\right)+\frac{\mu_{R}^2 m^2}{\lambda_{R}}
-\frac{m^4}{2\lambda_{R}} \; ,
\end{eqnarray}
in the limit $d=3$, where we used that
\begin{equation}
\left(\frac{m^2}{\Lambda^{2}}\right)^{(d-3)/2}\Gamma{\left(\frac{3-d}{2}
\right)}\sim{\frac{2}{(3-d)}}+\gamma-{\ln}\left(\frac{m^2}
{\Lambda^{2}}\right)+O(d-3) \; ,
\end{equation}
with $\gamma$ being the Euler constant. At this point we choose
the scale $\Lambda$ to be
\begin{equation}
\Lambda^{2}=\frac{\mu_{R}^{2}e^{-(\gamma-1/2)}}{4\pi} \; ,
\label{Lam}
\end{equation}
which leads to
\begin{equation}
V_{\text{eff}}(\varphi)=\frac{m^2}{2} {\varphi^{2}}+\frac{m^4}
{64\pi^{2}}\left[{\ln}\left(\frac{m^2}{\mu^{2}_{R}}\right)-
\frac{1}{2}\right]-\frac{(m^2-\mu^{2}_{R})^{2}}{2\lambda_{R}}+
\frac{\mu^{4}_{R}}{2\lambda_{R}} \; ,
\end{equation}
which is the standard result.

The renormalized expression for $m^2$ can be obtained either by
substituting the renormalization prescription into Eq.\
(\ref{stat:m}), or by minimizing the renormalized $V_{\text{eff}}$
with respect to $m^2$.
\begin{equation}
\frac{\partial V_{\text{eff}}(\varphi)}{\partial{m^2}}=0
\Rightarrow m^2 =\mu^{2}_{R}+\frac{\lambda_R}{2} \varphi^{2}
+\frac{\lambda_R}{32\pi^{2}} m^2 \;{\ln}\left(\frac{m^2}{\mu^{2}_{R}}
\right) \; .
\end{equation}

%%%%%%%%%%%%%%%%%%%%%%%%%%%%%%%%%%%%%%%%%%%%%%%%%%%%%%%%%%%%%%%%%%%%%%%%%

\section{Scalar field equations of motion in Robertson-Walker space-time}

In this section we obtain the renormalized equations of motion
for a self-interacting scalar field using a simplified Gaussian
approximation. We consider a flat Robertson-Walker space-time
with the line element
\begin{equation}
ds^2 = dt^2 - a^2(t) d{\bf x}^2 \; ,
\label{rw}
\end{equation}
where $a(t)$ is the scale factor. We assume minimal coupling between
gravity and the scalar field and that the scalar field dynamics is
governed by the Lagrange density
\begin{equation}
{\cal L} = a^d \left [
\frac{1}{2}~ g^{\mu\nu}~ \partial_\mu \Phi \partial_\nu \Phi
- U(\Phi) \right ] \; ,
\label{lag:dens}
\end{equation}
where the potential $U$ is given by Eq.\ (\ref{pot:u}). Although we
are mainly interested in physical space-time dimensionality, $d=3$, we
consider the theory in $d$ spatial dimensions in order to regularize
the theory in later discussions.

The canonical momentum $\Pi$ is defined by
\begin{equation}
\Pi \equiv \frac{\partial {\cal L}}{\partial \dot\Phi} = a^d \dot \Phi \; ,
\end{equation}
and the Hamiltonian density of the system is
\begin{equation}
{\cal H} = a^{d}\left\{\frac{1}{2} \left [ a^{-2d}\Pi^{2}
+a^{-2}({\bf \nabla}\Phi)^{2}\right ] +U(\Phi)\right\} \; .
\end{equation}

The use of the general Gaussian {\em Ansatz} (\ref{gauss:pure}) leads
to coupled integro-differential equations for the Fourier modes of the
kernels $\Omega$ and $\Sigma$ \cite{sep}. Therefore, we must solve an
infinity (large) number of coupled equations either analytically or
numerically when we apply this approximation to study a physical
problem.  In order to reduce the number of free parameters and
equations, we introduce a simplified Gaussian {\em Ansatz}, which is
obtained by fixing the functional dependence on ${\bf k}$ of the
kernels $\Omega$ and $\Sigma$. In this work, we assume that these
kernels, appearing in the Gaussian trial state (\ref{gauss:pure}),
have a form similar to their static ones, that is
\begin{equation}
\Omega({\bf k},t)=\frac{a^{1-d}}{2\sqrt{k^{2}+a^{2}\alpha(t)}} \;\;,
\label{omega}
\end{equation}
and
\begin{equation}
\Sigma({\bf k},t)=-\frac{a^{d-1}\beta}{8[k^{2}+a^{2}\alpha(t)]^{n}} \;\;,
\label{sigma}
\end{equation}
where $\alpha$ and $\beta$ are the variational parameters and $n$ is
conveniently chosen to control the infinities in the approximation.
Clearly, the above {\em Ansatz} recovers the vacuum solution
(\ref{stat}) in the static limit, provided we take $a=1$, $\beta=0$
and $\alpha = m^2$.

The effective action (\ref{dirac}) evaluated in the Gaussian trial
state (\ref{gauss:pure}) with the kernels (\ref{omega}) and
(\ref{sigma}) is
\begin{eqnarray}
\Gamma &=& \int dt \int_{{\bf x}}\left\{\hat{\pi}\dot{\varphi}-a^{d}\left[\frac
{1}{2}a^{-2d}\hat{\pi}^{2}+\frac{\mu^2}{2}\varphi^{2}
+\frac{\lambda}{4!}{\varphi}^{4}\right]-\frac{1}{16}(1-d) H\beta
\int_{{\bf k}}\frac{1}{(k^{2}+a^{2}\alpha)^{n+1/2}} \right. \nonumber\\
 &+& \frac{1}{32}a^2 \beta(2H\alpha+\dot{\alpha})
\int_{{\bf k}}\frac{1}{(k^{2}+a^{2}\alpha)^{n+3/2}}-\frac{1}{64}a^2\beta^{2}
\int_{{\bf k}}\frac{1}{(k^{2}+a^{2}\alpha)^{2n+1/2}}  \nonumber\\
 &-&\frac{1}{2}a^{-1}\int_{{\bf k}}\sqrt{k^{2}+a^{2}\alpha}+
\frac{1}{4}a\alpha \int_{{\bf k}}\frac{1}{\sqrt{k^{2}+a^{2}\alpha}}
-\frac{1}{4}a\left(\mu^{2}+\frac{\lambda}{2} {\varphi}^{2}\right)
\int_{{\bf k}}\frac{1}{\sqrt{k^{2}+a^{2}\alpha}}  \nonumber\\
 &+& \left. \frac{\lambda}{32} a^{2-d}\int_{{\bf k},{\bf k}'}\frac{1}
{\sqrt{k^{2}+a^{2}\alpha}}\frac{1}{\sqrt{k'^{2}+a^{2}\alpha}} \right\} \; ,
\end{eqnarray}
where $H= \dot{a}/a$ is the Hubble constant. In the limit $d=3$
infinities appear in $\Gamma$, due to the short distance behavior of
the kernels $\Omega$ and $\Sigma$. However, choosing $n>1$ we do not
introduce further infinities besides the one appearing in
$V_{\text{eff}}$\cite{aviso}.

Using dimensional regularization, we can separate the divergent
and finite parts of $\Gamma$ as
\begin{eqnarray}
\Gamma &=& \int dt \int_{{\bf x}}\frac{a^{d}}{(4\pi)^{(d+1)/2}(1-d)}
\left(\frac{\alpha}{\Lambda^{2}}\right)^{(d-3)/2}\left[\frac{(1-d)}{(1+d)}
\alpha^{2}+\left(\mu^{2} + \frac{\lambda}{2} \varphi^2 \right)\alpha
\right. \nonumber\\
 &+& \left. \frac{\lambda}{2}~ \frac{\alpha^{2}}{(4\pi)^{(d+1)/2}(1-d)}\left
(\frac{\alpha}{\Lambda^{2}}\right)^{(d-3)/2}\Gamma\left(\frac{3-d}{2}\right)
\right]+\text{terms finite at $d=3$} \; .
\end{eqnarray}
The divergent part of $\Gamma$ is similar to the divergences
encountered in $V_{\text{eff}}$, see Eq.\ (\ref{veff:0}). This allow
us to conclude that $\Gamma$ can be made finite in the limit $d=3$
using the renormalization prescription (\ref{mu:r})--(\ref{lambda:r}).
It is straightforward to verify that $\Gamma$ becomes finite in the
limit $d=3$ by this renormalization prescription:
\begin{eqnarray}
  \Gamma &=& \int dt \int_{{\bf x}}\left\{\hat{\pi}\dot{\varphi}-\frac{a^{-3}}
  {2}\hat{\pi}^{2}+\frac{1}{8} H\beta I_{n+1/2} +\frac{1}{32}a^{2}\beta(2H
  \alpha+\dot{\alpha})I_{n+3/2}-\frac{1}{64}a^{-1}\beta^{2}I_{2n+1/2} \right.
\nonumber\\
   &-& \left. a^{3} \left[
       \frac{\alpha}{2}\varphi^{2} + \frac{\alpha^{2}}{64\pi^{2}}
       \left(\ln\left(\frac{\alpha} {\mu_{R}^{2}}\right)-\frac{1}{2}\right)
       -\frac{(\alpha-\mu_{R}^{2})^{2}}{2\lambda_{R}}+\frac{\mu_{R}^{4}}
       {2\lambda_{R}}\right] \right\} \; ,
\end{eqnarray}
where we chose the scale $\Lambda$ as in Eq.\ (\ref{Lam}) and we
defined the integrals
\begin{equation}
I_{j}\equiv \int_{{\bf k}}\frac{1}{(k^{2}+a^{2}\alpha)^{j}}=\frac{1}{2^{d}\pi
^{d/2}}(a^{2}\alpha)^{(d-2j)/2}\frac{\Gamma(j-d/2)}{\Gamma(j)} \; .
\end{equation}

At this point it is interesting to compare our results with the ones
for general Gaussian {\em Ansatz} as shown in Ref.\ \cite{r:td}.  In
both approximations the dynamical equations of motion become finite by
the vacuum sector renormalization prescription. Moreover, the large
${\bf k}$ behavior of the kernels in these two approximations are
similar: in the general Gaussian {\em Ansatz} it is required that
$\Omega \simeq O(k^{-1})$ and $\dot\Omega \simeq O(k^{-3})$, while in
our approximation $\Omega$ exhibits the same high energy behavior, by
construction, and $n > 1$, that means that the asymptotic ${\bf k}$
dependence of $\Sigma$ is similar in both approximations.

By varying the renormalized effective action $\Gamma$ with respect to
the parameters $\varphi$, $\hat\pi$, $\alpha$, and $\beta$, we obtain
four coupled variational equations
\begin{eqnarray}
\frac{\delta \Gamma}{\delta \varphi}=0 &\Rightarrow&
 \dot{\hat{\pi}}=-a^{3}\alpha \varphi \;,
\label{eqm:1}
\\
\frac{\delta \Gamma}{\delta \hat{\pi}}=0 &\Rightarrow&
\dot{\varphi}=a^{-3}\hat{\pi} \;,
\\
\frac{\delta \Gamma}{\delta \beta}=0 &\Rightarrow&
\dot{\alpha}=  a^{-3}\beta \frac{I_{2n+1/2}}{I_{n+3/2}}
-6 ~\frac{n}{n-1} H \alpha \; ,
\\
\frac{\delta \Gamma}{\delta \alpha} = 0 &\Rightarrow& \dot{\beta} = a^{-2}
I_{n+3/2}^{-1}\left\{2 \frac{(4n+1)~(3n-1)}{2n-1} a \beta^{2}
I_{2n+3/2} \right. \nonumber\\
 &-& \left. 32a^{3}\left[ \frac{\varphi^{2}}{2}+\frac{\alpha}
{32\pi^{2}}~\ln\left(\frac{\alpha}{\mu_{R}^{2}}\right)- \frac{(\alpha-
\mu_{R}^{2})}{\lambda_{R}}\right] \right\}-2~\frac{5n^2-n-1}{n-1}H \beta \; .
\label{eqm:2}
\end{eqnarray}
Due to the choice of our trial state, the equations of motion for the
parameters $\varphi$ and $\hat\pi$ are free field ones with a
time-dependent mass $\alpha$, whose dynamics is dictated by the last
two equations \cite{turk}.

%%%%%%%%%%%%%%%%%%%%%%%%%%%%%%%%%%%%%%%%%%%%%%%%%%%%%%%%%%%%%%%%%%%%%%

\section{Renormalizing the energy-momentum tensor}

In order to write the semi-classical Einstein equation we must study
the renormalization of the expectation value of the energy-momentum
tensor in our trial state. The energy-momentum tensor for the scalar
field described by the Lagrange density (\ref{lag:dens}) is
\begin{eqnarray}
T_{\mu \nu}&\equiv&\frac{2}{\sqrt{-g}}\frac{\delta I}{\delta g^{\mu \nu}}
\nonumber\\
 &=& \partial_{\mu}\Phi \partial_{\nu}\Phi-g_{\mu \nu}\left[\frac{1}{2}
g^{\alpha \beta}\partial_{\alpha}\Phi \partial_{\beta}\Phi-\frac{\mu^2}{2}
\Phi^{2}-\frac{\lambda}{4!} \Phi^{4}\right] \; ,
\label{tmunu}
\end{eqnarray}
where $I$ is the action, $G_{\mu\nu} \equiv R_{\mu\nu} - \frac{1}{2}
g_{\mu\nu} R$ is the Einstein tensor and the notation $;\mu$ denotes
the covariant derivative with respect to the space-time index $\mu$.
In the functional Schr\"odinger picture, we express the
energy-momentum tensor operator in terms of the field operator
$\Phi({\bf x})$ and its canonical momentum $\Pi({\bf x})$ and evaluate
the expectation value in a given state.

In the flat Robertson-Walker metric (\ref{rw}), the expectation value
of $T_{\mu\nu}$ in the translationally invariant Gaussian state
(\ref{gauss:pure}) has the form
\begin{eqnarray}
 \langle T_{00} \rangle &=& \frac{a^{-2d}}{2}\hat{\pi}^{2}+\frac{\mu^2}{2}
 \varphi^{2}+\frac{\lambda}{4!} {\varphi}^{4}
 + \frac{a^{-2d}}{8}\Omega^{-1}({\bf x},{\bf x},t)+2a^{-d}(\Sigma{\Omega}
 \Sigma)({\bf x},{\bf x},t)
\nonumber\\
 &+& \left. \frac{1}{2}\left[-a^{-2}{\bf \nabla}_{{\bf x}}^{2}+\mu^{2}+
 \frac{\lambda}{2}
 \varphi^{2}+\frac{\lambda}{2}{\Omega({\bf x},{\bf x},t)}\right]
 \Omega({\bf x},{\bf y},t)\right |_{{\bf x}={\bf y}}
\nonumber\\
 &-& \frac{\lambda}{8} \Omega({\bf x},{\bf x},t) \Omega({\bf x},{\bf x},t)
\; ,
\label{tmunu:1}
\\
&& \nonumber \\
  \langle T_{ij} \rangle &=& a^{2}\delta_{ij}\left\{\frac{a^{-2d}}{2}\hat{\pi}
  ^{2}-\frac{\mu^2}{2} \varphi^{2}-\frac{\lambda}{4!} {\varphi}^{4}
  \right.
  +\frac{a^{-2d}}{8}\Omega^{-1}({\bf x},{\bf x},t)+2a^{-d}(\Sigma{\Omega}
  \Sigma)({\bf x},{\bf x},t)
\nonumber\\
  &-& \left. \frac{1}{2}\left[-\left(1-\frac{2}{d}
   \right)a^{-2}{\bf \nabla}_{{\bf x}}
  ^{2}+\mu^{2}+\frac{\lambda}{2} {\varphi}^{2}+\frac{\lambda}{2} {\Omega}
  ({\bf x},{\bf x},t)\right]\Omega({\bf x},{\bf y},t)\right |_{{\bf x}={\bf y}}
\nonumber\\
  &+& \left. \frac{\lambda}{8}{\Omega}({\bf x},{\bf x},t)\Omega({\bf x},
  {\bf x},t) \right\}
\; ,
\label{tij}
\\
&& \nonumber \\
 \langle T_{0i} \rangle &=& 0 \; .
\label{tmunu:2}
\end{eqnarray}

As expected, the energy-momentum tensor matrix element is diagonal
and can be expressed in terms of the average energy density $\langle
\epsilon\rangle$ and pressure $\langle p \rangle$. In four space-time
dimensions, we have
\begin{equation}
\langle T_{\mu\nu}\rangle = \text{diag} \left (
\langle \epsilon \rangle, a^2 \langle p \rangle,
 a^2 \langle p \rangle,  a^2 \langle p \rangle \right ) \; .
\end{equation}
Once more we can witness from (\ref{tmunu:1}-\ref{tmunu:2}) that
infinities may appear in the limit $d=3$ because of the short distance
behavior of the kernels $\Omega$ and $\Sigma$.

Now we discuss how to obtain the finite, renormalized expectation
value of $T_{\mu\nu}$ in our Gaussian {\em Ansatz}, given by Eqs.\
(\ref{gauss:pure}), (\ref{omega}), and (\ref{sigma}). First of all, we
evaluate the dimensionally regularized expression for $\langle
T_{\mu\nu} \rangle$, which is finite and well behaved.  A nice feature
of this regularization procedure is that it preserves the general
covariance of $\langle T_{\mu\nu} \rangle$.  Substituting
(\ref{omega}) and (\ref{sigma}) into (\ref{tmunu:1}-\ref{tij}), we
obtain
\begin{eqnarray}
\langle T_{00} \rangle
   &=& \frac{a^{-2d}}{2}\hat{\pi}^{2}+\frac{\mu^2}{2} \varphi^{2}
    +\frac{\lambda}{4!} {\varphi}^{4}
    +\frac{1}{64}a^{-d-1}\beta^{2}I_{2n+1/2}
\nonumber\\
   &+& \frac{1}{(4\pi)^{(d+1)/2}(1-d)}\left(\frac{\alpha}{\Lambda^{2}}\right)
    ^{(d-3)/2}\Gamma \left(\frac{3-d}{2}
    \right)\left[\frac{2}{(1+d)}\alpha^{2}-\alpha^{2}+\left(\mu^{2}
    +\frac{\lambda}{2} \varphi^{2}\right)\alpha \right.
\nonumber\\
   &+& \left. \frac{\lambda}{2} \frac{\alpha^{2} }{(4\pi)^{(d+1)/2}(1-d)}
    \left(\frac{\alpha}{\Lambda^{2}}\right)^{(d-3)/2}\Gamma
    \left(\frac{3-d}{2}\right)\right] \; ,
\\
&& \nonumber \\
\langle T_{ij} \rangle
   &=& a^{2}\delta_{ij}\left\{\frac{a^{-2d}}{2}\hat{\pi}^{2}+\frac{\mu^2}{2}
    \varphi^{2}+\frac{\lambda}{4!} {\varphi}^{4}+\frac{1}{64}a^{-d-1}
    \beta^{2}I_{2n+1/2} \right.
\nonumber\\
   &+& \frac{1}{(4\pi)^{(d+1)/2}(1-d)}\left(\frac{\alpha}{\Lambda^{2}}\right)
    ^{(d-3)/2}\Gamma \left(\frac{3-d}{2}
    \right)\left[\frac{2}{d(1+d)}\alpha^{2}+\left(1-\frac{2}{d}\right)
    \alpha^{2} \right.
\nonumber\\
   &-& \left. \left.\left(\mu^{2}+\frac{\lambda}{2} \varphi^{2}\right)\alpha
    -\frac{\lambda}{2} \frac{\alpha^{2}}{(4\pi)^{(d+1)/2}(1-d)}
    \left(\frac{\alpha}{\Lambda^{2}}\right)^{(d-3)/2}\Gamma \left(\frac{3-d}
    {2}\right)\right] \right\} \; .
\end{eqnarray}

Next, we express the expectation value of the energy-momentum tensor
in terms of the renormalized parameters $\mu_R$ and $\lambda_R$. In
order to do so, we substitute Eqs.\ (\ref{mu:r}) and (\ref{lambda:r})
into the last two expressions, resulting in
\begin{eqnarray}
\langle T_{00} \rangle &=& \frac{a^{-6}}{2}\hat{\pi}^{2}+\frac{1}{64}a^{-4}
                           \beta^{2}I_{2n+1/2} + \frac{\alpha}{2}\varphi^{2}
\nonumber\\
                       &+& \frac{\alpha^2}{64\pi^{2}}
                           \left[\ln\left(\frac{\alpha}{\mu_{R}^{2}}
                           \right)-\frac{1}{2}\right]
                           -\frac{(\alpha-\mu_{R}^{2})^{2}}{2\lambda_{R}}
                           +\frac{\mu_{R}^{4}}{2\lambda_{R}} \; ,
\\
&& \nonumber \\
\langle T_{ij} \rangle &=& a^{2}\delta_{ij}\left\{\frac{a^{-6}}{2}
    \hat{\pi}^2 + \frac{1}{64} a^{-4}\beta^{2}I_{2n+1/2}
    - \frac{\alpha}{2}\varphi^{2} \right.
\nonumber\\
                       &-& \left. \frac{\alpha^2}{64\pi^{2}}
                           \left[\ln\left(\frac{\alpha}{\mu_{R}^{2}}
                           \right)-\frac{1}{2}\right]+\frac{(\alpha
                           -\mu_{R}^{2})^{2}}{2\lambda_{R}}
                           -\frac{\mu_{R}^{4}}{2\lambda_{R}}
\right\} \; ,
\end{eqnarray}
where we took the limit $d \rightarrow3$ and chose the scale $\Lambda$
according to Eq.\ (\ref{Lam}).

One important feature of our Gaussian approximation is that the
expectation value of $T_{\mu\nu}$ turns out to be finite once we take
into account the mass and coupling constant renormalization, and that
$n>1$. Therefore, the presence of interactions, more specifically the
term $-g_{\mu\nu} (\lambda/8) \Omega \Omega$, leads to the cancelation
of the infinities which exist in the free scalar field model
\cite{sep}.

Now we discuss the limit $\lambda_R = 0$. In this limit, the
renormalization prescription tells us that
\begin{equation}
\lambda = 0 \;\; ; \;\; \mu^2_R = \mu^2 \; ,
\end{equation}
and hence we get back the unrenormalized free theory result for
$\langle T_{\mu\nu} \rangle$ in which the divergences reappear from
the $1/\lambda_R$ terms. In order to have a well behaved free theory
limit we must subtract the terms that diverge in the limit
$\lambda_R=0$. Moreover, this subtraction may be justified as a
renormalization of coupling constants in a generalized Einstein
equation provided that the entire subtraction is expressible in terms
of covariantly conserved tensors \cite{sep}.

In the free theory limit, the contribution $- g_{\mu\nu} (\alpha
-\mu_{R}^{2})^2/2\lambda_R$ vanishes since the structure equation of
motion for $\alpha$ leads to $\alpha=\mu^2$ for $\lambda_R=0$.
Therefore, we define the renormalized expectation value of the
energy-momentum tensor as
\begin{equation}
\langle T_{\mu\nu} \rangle_R \equiv  \langle T_{\mu\nu} \rangle -
g_{\mu\nu} \frac{\mu_R^4}{2\lambda_R} \; .
\end{equation}
Notice that this subtraction is basically a redefinition of the
cosmological constant.

%%%%%%%%%%%%%%%%%%%%%%%%%%%%%%%%%%%%%%%%%%%%%%%%%%%%%%%%%%%%%%%%%%%%%%

\section{Conclusions and discussion}

In this work we used Dirac's variational principle and a simple
Gaussian {\em Ansatz} to describe the time evolution of a
self-interaction scalar field in flat Robertson-Walker space-time.
Unlike ordinary perturbation theory, this approximation reflects some
of the non-linear features of the full quantum field theory which may
be important for understanding various physical processes.  Our trial
wave functional was obtained from a general Gaussian state by choosing
the momentum dependence of its kernels in such a way that we can
recover the conventional vacuum in the limit that we have a Minkowski
space-time. Nevertheless, we should point out that our vacuum state
for a Robertson-Walker space-time is orthogonal to the one obtained in
Refs.\ \cite{r:td,sep}, where the Gaussian adiabatic vacuum was used,
since they differ by terms of order $O(k^{-3})$ for $k \rightarrow
\infty$.

In principle, the equations for the variational parameters and the
semi-classical Einstein equation may be used to study dynamical
question about the universe, such as stability of the de Sitter space
and the conditions for inflation setting in.  In this case, however,
we are confined to the chaotic inflation scenario since our {\em
Ansatz} does not describe correctly the dynamics of the low momentum
modes, as well as the Gaussian approximation in higher-dimensional
field theory suffers from well-known shortcomings, analogous to the
ones that appear in the the large-$N$ approximation.

%**********
\acknowledgments

One of us (O.J.P.E.) would like to thank the kind hospitality of the
Institute for Elementary Particle Research, University of
Wisconsin--Madison, where the final part of this work was done.  This
work was partially supported by the U.S.\ Department of Energy under
Grants Nos.\ DE-FG02-95ER40896 and DE-FG02-91ER40661, by the
University of Wisconsin Research Committee with funds granted by the
Wisconsin Alumni Research Foundation, by Conselho Nacional de
Desenvolvimento Cient\'{\i}fico e Tecnol\'ogico (CNPq), by
Coordenadoria de Aperfei\c{c}oamento de Pessoal de Ensino Superior
(CAPES), and by Funda\c{c}\~ao de Amparo \`a Pesquisa do Estado de
S\~ao Paulo (FAPESP).

%**********
% REFERENCES
%**********


\begin{references}

\bibitem[*]{hcr} Electronic address: hreis@if.usp.br (internet)
or 47602::hreis (decnet).

\bibitem[\dag]{ojpe} Electronic address: eboli@phenom.physics.wisc.edu
(internet) or phenoa::eboli (decnet). On leave of absence from
Instituto de F\'{\i}sica, Universidade de S\~ao Paulo, C.P.\ 66318,
05389-970 S\~ao Paulo, Brazil.


\bibitem{inf:1} For a review see L.\ Abbott and S.-Y.\ Pi, {\em
    Inflationary Cosmology}, World Scientific, Singapore, 1986; \\
    A.\ Linde, {\em Particle Physics and Inflationary Cosmology},
    Harwood Academic Publishers, Chur, Switzerland, 1990.

\bibitem{str} See, for instance, A.\ Vilenkin, Phys.\ Rep.\ {\bf 121},
  263 (1985); \\
   R.\ Brandenberger, in {\em Field Theory and Particle
    Physics}, VII Jorge Andr\'e Swieca Summer School, Campos do
  Jord\~ao, Brazil, 1993, edited by O.\ J.\ P.\ \'Eboli and V.\ O.\
  Rivelles (World Scientific, Singapore, 1994).

\bibitem{sch} For a review, see R.\ Jackiw and S.-Y.\ Pi contributions
 to {\em Field Theory and Particle Physics}, V Jorge Andr\'e Swieca
 Summer School, Campos do Jord\~ao, Brazil, 1989, edited by O.\ J.\
 P.\ \'Eboli, M.\ Gomes, and A.\ Santoro (World Scientific, Singapore,
 1990).

\bibitem{r:sta} K.\ Symanzik, Nucl.\ Phys.\ {\bf B190}, 1 (1981); \\
M.\ L\"uscher, Nucl.\ Phys.\ {\bf B254}, 52 (1985).

\bibitem{r:td} S.-Y.\ Pi and M.\ Samiullah, Phys.\ Rev.\ {\bf D36},
3128 (1987); \\
F.\ Cooper and E.\ Mottola, Phys.\ Rev.\ {\bf D36}, 3114 (1987).

\bibitem{ejp} O.\ J.\ P.\ \'Eboli, R.\ Jackiw, and S.-Y.\ Pi,
Phys.\ Rev.\ {\bf D37}, 3557 (1988).

\bibitem{jk} R.\ Jackiw and A.\ Kerman, Phys.\ Lett.\ {\bf A71}, 158
(1979); \\
R.\ Jackiw, Int.\ J.\ Quantum Chem.\ {\bf 17}, 41 (1980).

\bibitem{bv} R.\ Balian and M.\ V\'en\'eroni, Phys.\ Rev.\ Lett.\
{\bf 47}, 1353 (1981); {\bf 47}, 1765(E) (1981);\\
Ann.\ Phys.\ {\bf 164}, 334 (1985).

\bibitem{semi} See, for instance, A.\ Albrecht and R.\ Brandenberger,
Phys.\ Rev.\ {\bf D31}, 1225 (1985); \\
A.\ Albrecht, R.\ Brandenberger, and R.\ Matzner, {\em ibid.}
{\bf 32}, 1280 (1985); {\bf 35}, 429 (1987); \\
H.\ Kurki-Suoni, J.\ Centrella, R.\ Matzner, and
J.\ Wilson, {\em ibid.} {\bf 35}, 435 (1987); \\
R.\ Brandenberger, H.\ Feldman, and J.\ MacGibbon, {\em ibid.} {\bf 37},
2071 (1988).

\bibitem{line} A.\ H.\ Guth and S.-Y.\ Pi, Phys.\ Rev.\ {\bf D32},
1899 (1985).

\bibitem{sep} O.\ J.\ P.\ \'Eboli, S.-Y.\ Pi, and M.\ Samiullah, Ann.\ Phys.\
{\bf 193}, 102 (1989); \\
M.\ Samiullah, O.\ J.\ P.\ \'Eboli, and S.-Y.\ Pi, Phys.\ Rev.\ {\bf D44},
2335 (1991).

\bibitem{boya} D. Boyanovsky and Da-Shin Lee, Phys.\ Rev.\
{\bf D48}, 800 (1993); \\
 D. Boyanovsky, H.\ J.\ de Vega, and R.\ Holman, {\em ibid.}
{\bf 49}, 2769 (1994).

\bibitem{dirac} P.\ A.\ M.\ Dirac, Proc.\ Cambridge Philos.\ Soc.\
{\bf 26}, 376 (1930).

\bibitem{jj} C.\ G.\ Bollini and J.\ J.\ Giambiagi, Phys.\ Lett.\ {\bf 40B},
566 (1972); \\
G.\ 't Hooft and M.\ Veltman, Nucl.\ Phys.\ {\bf B44}, 189 (1972); \\
J.\ F.\ Ashmore, Nuovo Cimento Lett.\ {\bf 4}, 289 (1972).

\bibitem{presc} S.\ Coleman, R.\ Jackiw, and D.\ Politzer, Phys.\ Rev.\
{\bf D10}, 2491 (1974); \\
W.\ Bardeen and M.\ Moshe, {\em ibid.} {\bf28}, 1372 (1983).

\bibitem{aviso} For a Minkowski space-time, we can also choose $n=1$
since the divergent terms for $n=1$ are multiplied by the Hubble
constant $H$.

\bibitem{turk} For a Minkowski space-time ($a=1$), our results agree
with the ones obtained by S. Turkoz, Ph.D. thesis, Massachusetts
Institute of Technology, 1989.


\end{references}
\end{document}